\documentclass[prl,twocolumn,superscriptaddress]{revtex4}
\usepackage{graphicx}
\begin{document}

\title{Theory for Gossamer and Resonating Valence Bond Superconductivity}
\author{J. Y. Gan$^{1}$, F. C. Zhang$^{2,3}$, Z. B. Su}
\affiliation{Institute of Theoretical Physics, Chinese Academy of Sciences, Beijing, China\\
$^2$ Department of Physics, The University of Hong Kong, Hong Kong\\
$^3$ Department of Physics, University of Cincinnati, Cincinnati, Ohio 45221~\cite{cinti} }

\begin{abstract}
We use an effective Hamiltonian for two-dimensional Hubbard model including an antiferromagnetic spin-spin coupling
term to study recently proposed gossamer superconductivity.  We formulate a renormalized mean field theory
to approximately take into account the strong correlation effect in the partially projected Gutzwiller
wavefucntions.  At the half filled, there is a first order phase transition to separate a Mott insulator
at large Coulomb repulsion $U$ from a gossamer superconductor at small $U$. Away from the half filled,
the Mott insulator is evolved into an resonating valence bond state, which is adiabatically connected
to the gossamer superconductor.
\end{abstract}

\pacs{75.10.Jm, 05.30.Pr,75.10.-b}

\maketitle

\section*{\bf 1. Introduction}
Since the discovery of high temperature superconductivity in the cuprates~\cite{bednorz,chu},
there have been a lot of theoretical efforts trying to understand the microscopic mechanism for it.
One of the scenarios was
initiated by Anderson~\cite{anderson}, who proposed the idea of resonating valence
bond (RVB) state for the observed unusual properties in these compounds. A minimum model
for cuprates was argued to be 2-dimensional Hubbard or its equivalent $t-J$ model in the
large U limit~\cite{anderson,zhang-rice}.
In the RVB picture,
each lattice site is either unoccupied or singly occupied by a spin-up or spin-down electron.
The spins are coupled antiferromagnetically without long range order. The charge carriers move in the
spin background and condense to a superconducting state~\cite{baskaran87,zhang88,gros,kotliar,fukuyama,
palee,randeria}. In this scenario, the undoped cuprate with density of one electron per site
is a Mott insulator, and the superconductor is viewed as a doped Mott insulator.
Many experimentally observed properties in cuprates, such as the d-wave symmetry
in superconductivity~\cite{harlingen,tsuei} ,the pseudogap phenomena~\cite{pseudogap}, and
the linear doping dependence of the superfluid density in the underdoped region~\cite{uemura},
seem to be consistent with the RVB mean field theory.  On the other hand, while
mean field theories and variational calculations show the superconductivity in the doped
Hubbard or $t-J$ models, more direct numerical calculations on these models
remain controversial and have been unable to provide unambiguous answers to this
question~\cite{sorrela,tklee,assaad,gubernatis,white,jarrell}.

Very recently, Laughlin has proposed an interesting new notion, the gossamer superconductivity,
for high T$_c$ superconducting Cu-oxides~\cite{laughlin}.  In a gossamer superconductor,
the superfluid density is tenuous, in contrast to the conventional superconductor. He proposed
an explicit many body wavefunction for that state, which is a partially Gutzwiller projected
BCS state (Eq. (2) below). The partial projection operator
enables one
to construct its inverse operator. Using these operators, Laughlin has further proposed
a Hamiltonian, for which the partially projected BCS state is an exact ground state.
Exact solutions play important role in many physical problems. By expanding that Hamiltonian,
Laughlin showed that the superconducting ground state requires a large attractive interaction
in addition to a large on-site Coulomb repulsion. This raises the question if the Hubbard or $t-J$
models capture the basic physics in cuprates.

In a previous paper, one of us argued that the effective Hamiltonian of the Hubbard model
acting on the Gutzwiller's wavefunction should include a spin-spin coupling term, and
the on-site Coulomb repulsion plays both the roles in projecting out (partially or completely)
the double occupied state and in generating an attractive pairing interaction~\cite{zhang02}.
Although any variational calculation cannot make a conclusion about the exact ground state,
it is clear that the spin-spin coupling generated from the Hubbard $U$ should capture some
basic features in cuprates including its superconductivity. Nevertheless,
Laughlin's idea of gossamer superconductivity is interesting. The wavefunction he proposed
enables us to study the phase transition between a Mott insulator and a gossamer superconductor
at the half filled electron density and to study the strong coupling RVB state from a new
viewpoint: namely its adiabatic continuation to the intermediate coupling gossamer state.

In this paper, we use an effective Hamiltonian (Eq. (1) below)
for 2-dimensional Hubbard in square lattice to systematically
study the partially projected Gutzwiller wavefunction.  We are interested in
the competition between the Mott insulator and the superconductor.  Here we shall neglect the
possible antiferromagnetism in the model, which will be a subject in a future
publication.  We use Gutzwiller's approximation
to replace the strong correlation in the projection by a set of renormalized factors, and to
use a renormalized mean field theory to study the ground state and the elementary excited
states of the system.  Our main results can be summarized below.  At the half filling,
the ground state is a Mott insulator at large $U$, and a gossamer superconductor at small
$U$. The transition is first type in the physically interesting parameter region. The
charge carrier density and the superconducting order parameter change discontinuously
from zero in the Mott insulating phase to a finite value at the critical value of $U$.
Away from the half filled, the gossamer superconducting state changes continuously from
its state at the half filled, while the Mott insulating phase becomes RVB superconducting.
The gossamer and RVB suerconducting states have the same pairing symmetry, and their
superconducting order parameters are both suppressed by a unified renormalized factor, which
quantitatively characterizes the smallness of the superfluid density.
Therefore, the gossamer and RVB
superconducting states are adiabatically connected. The gossamer superconducting state
at the half filled
may be viewed as a RVB state with equal number of independent empty and doubly occupied sites.
These empty and doubly occupied sites provide "parking space" for singly occupied electrons
to move through the lattice. From this point of view, the relative reduction of $U$ respected to the kinetic
energy, which may be realized by applying the pressure~\cite{baskaran}, plays a similar role
as the chemical doping.  The gossamer superconductivity may have already been
realized in organic superconductors~\cite{organic}.
In Fig. 1, we schematically show snap shots for a Mott insulator, an RVB state,
and a gossamer superconducting state.

\begin{figure}
\includegraphics*[width=3in,height=1in,angle=0]{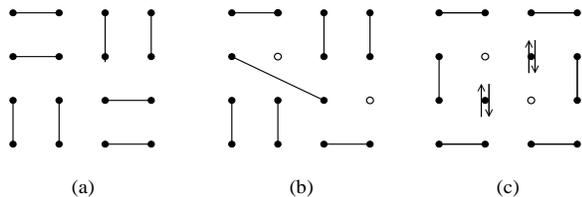}
\caption{Snap shots for a) A spin liquid Mott insulator; b). An RVB superconducting state;
and c). A gossamer superconducting state.Each bond connects an up-spin and a down-spin and
open circle respresents the hole.}
\label{fig:pd}
\end{figure}

This paper is organized as follows.  In Section 2, we introduce the model and the variational wavefunction.
In Section 3, we formulate a renormalized mean field theory to study the variational wavefunction.
Section 4 is devoted to the phase transition between the Mott insulator and  the gossamer superconductor
at the half filled. Detailed discussions on the gossamer and RVB superconductivity are given in Section 5.
The paper is concluded with a summary in Section 6.

\bigskip
\section*{\bf 2. The Model and the Variational Wavefunction}

We study an effective Hubbard Hamiltonian in a square lattice,
\begin{eqnarray}
H & = & H_t + H_s + H_U \\
H_t & = & - \sum\limits_{<ij>\sigma}(t_{ij}c_{i\sigma}^\dagger c_{j\sigma} + h.c.) \nonumber\\
H_s & = & J \sum\limits_{<ij>}\vec{S}_i\cdot\vec{S}_j \nonumber\\
H_U & = & U\sum\limits_i n_{i\uparrow}n_{i\downarrow} \nonumber
\end{eqnarray}
In the above equations, $c_{i\sigma}$ is the annihilation operator of an electron of spin $\sigma$
at the lattice site $i$, and $n_{i\sigma}=c_{i\sigma}^\dagger c_{i\sigma}$.
The sum is over the nearest neighbor pairs of $<ij>$,
and $U > 0$ is the intra-site Coulomb repulsion. Without loss of generality, we consider the case $t >0$.
In this Hamiltonian, we have introduced an antiferromagnetic spin-spin coupling
term to account for the effect of the virtual electron hopping process.
In the large $U$ limit, $J\approx 4t^2/U$. This model may be viewed
as an effective Hamiltonian of the Hubbard model.
The inclusion of the antiferromagnetic
spin coupling appears consistent with the weak coupling renormalization group analysis~\cite{carsten},
and is appropriate in the variational approach studied here. In the limit $U \rightarrow \infty$,
the model is reduced to the $t-J$ model.
Very recently, a similar
form of the Hamiltonian has been derived by using two subsequent
canonical transformations starting from the Hubbard model at large $U$ limit~\cite{yue}.
In the Hilbert space with the fixed number of the doubly occupied electron sites on the lattice,
it has been shown that the Hubbard model may be mapped onto Eq. (1)
with the constraint that the hopping process in $H_t$ is limited to the corresponding
Hilbert space, namely it does not change the electron double occupation.
Here we shall consider Eq.(1) from a phenomenological point of view,
and study its solutions within the framework of Gutzwiller's variational approach. 
We shall consider $J$ to be an independent parameter. 

Due to the perfect nesting and the van Hove singularity in 
the density of state, the ground state of Hamiltonian (1) at the half filling
(electron density $n =1$ per site)
is an
antiferromagnet for arbitrarily small value of $U$ even in the absence of the spin-spin
coupling term.  The spin-spin coupling further enhances the magnetism.
In this paper, however, we shall focus on the insulating and metallic phases of the problem, and will
not include the magnetic long range order.

We study the model using 
a variational trial wavefunction proposed by Laughlin~\cite{laughlin},
\begin{eqnarray}
|\Psi_{GS} \rangle &=& \Pi_{\alpha}|\Psi_{BCS} \rangle \\
 \Pi_{\alpha}  &=&   \prod\limits_i (1 - \alpha n_{i\uparrow}n_{i\downarrow})
\end{eqnarray}
with $|\Psi_{BCS} \rangle$ a BCS-type superconducting state, given by
\begin{eqnarray}
|\Psi_{BCS} \rangle = \prod\limits_{\vec{k}} (u_{\vec{k}}
+ v_{\vec{k}}c_{\vec{k}\uparrow}^\dagger c_{-\vec{k}\downarrow}^\dagger) |0 \rangle.
\end{eqnarray}
where $ |0 \rangle$ is the vacuum, and $u_{\vec k}$ and $v_{\vec k}$ 
are variational parameters, satisfying the condition
\begin{eqnarray}
|u_{\vec k}|^2 + |v_{\vec k}|^2 = 1. \nonumber
\end{eqnarray}
$\Pi_{\alpha}$
is a projection operator to partially project out 
the doubly occupied electron states on
each lattice site $i$.  The state
$|\Psi_{GS} \rangle$ may be considered as a generalization of the previously studied
partially projected 
non-interacting electron state~\cite{gutzwiller,brinkman,vallhardt}
to include the superconducting state. In the limiting case
$u_{\vec{k}}v_{\vec{k}}=0$, $|\Psi_{BCS} \rangle$ is reduced to the non-interacting electron state,
and 
\begin{eqnarray}
|\Psi_{GS} \rangle \rightarrow   \Pi_{\alpha} |\Psi_{FL} \rangle, \nonumber
\end{eqnarray}
where $|\Psi_{FL} \rangle$ is the ground state of the non-interacting electron system, given by
$|\Psi_{FL} \rangle = \prod\limits_{\vec k, \sigma}c^{\dagger}_{\vec k\sigma}
c_{\vec k\sigma} |0 \rangle$, 
with the product running over all the $\vec k$'s within the Fermi
surface.  $|\Psi_{GS} \rangle$ is a natural generalization of the usual 
BCS state to strongly correlated systems. It connects the usual BCS state to the 
RVB state, characterized by the parameter  $\alpha$, which takes the value between 0 and 1.  
$\alpha=0$ is a normal BCS state.  At $\alpha=1$,  the projection operator projects out all the
doubly occupied electron states, and $|\Psi_{GS} \rangle$ is reduced to the
RVB state~\cite{anderson}. At the half filling and at $\alpha=1$, 
each lattice site is occupied by a single electron, and the system is a Mott insulator.
Therefore, the wavefunction $|\Psi_{GS}\rangle$ can be used to study superconductor-insulator transition.

\section*{\bf 3. The Renormalized Mean Field Theory}

We now proceed the variational calculations to determine the parameters $\alpha$ 
and $u_{\vec k}$,  $v_{\vec k}$.  Without loss of generality, we consider
the electron density $n \leq 1$.
The variational energy per site $ E = \langle H \rangle$ is given by
\begin{eqnarray}
E = Ud + \langle H_t \rangle + \langle H_J \rangle
\end{eqnarray}
where 
$d=\langle n_{i\uparrow}n_{i\downarrow} \rangle$ is the average electron double 
occupation number. 
$\langle Q \rangle$ is the expectation value of the operator $Q$ 
in the state $|\Psi_{GS} \rangle$.  For briefness, $\langle H_t \rangle, \langle H_J \rangle$ 
stand for their average values per site. 
$d$ is a function of $\alpha$, and $0\leq d \leq 1/4$.  
The first term in Eq. (5) is the intra-site Coulomb interaction energy,
while the second and the
third terms are the average kinetic and spin-spin correlation energies, respectively.

The variational calculations can be carried out using variational 
Monte Carlo method~\cite{gros,gros2,yokoyama,ogata}.  Here we use the
renormalized Hamiltonian approach to treat the projection operator approximately
~\cite{zhang88}.
In this approach, the effect of the projection operator is taken into account by
a classical statistical weighting  factor, which multiplies the quantum coherent
result of the non-projected state.  This method (Gutzwiller method hereafter)
was first proposed by 
Gutzwiller~\cite{gutzwiller},
and has been applied to study strongly correlated systems by many others
~\cite{brinkman,vallhardt,zhang88}. Let $ \langle Q  \rangle_0$ be the expectation value of $Q$
in the state $|\Psi_{BCS} \rangle$, then the hopping energy and the spin-spin corerlation
in the state $|\Psi_{GS} \rangle$ are related to those in the state $|\Psi_{BCS} \rangle$ by,
\begin{eqnarray}
\langle c_{i\sigma}^\dagger c_{j\sigma} \rangle = g_t \langle c_{i\sigma}^\dagger c_{j\sigma} \rangle_0 \nonumber\\
\langle \vec{S}_i\cdot\vec{S}_j \rangle = g_s \langle \vec{S}_i\cdot\vec{S}_j \rangle_0
\end{eqnarray}
The renormalized factors $g_t$ and $g_s$ are determined by the ratio of the probability
of the physical processes in the states $|\Psi_{GS} \rangle$ and $|\Psi_{BCS} \rangle$.
Following the counting method described in the literature~\cite{zhang88}, we have
\begin{eqnarray}
g_t &=& \frac{(n-2d)(\sqrt{d} + \sqrt{1-n+d})^2}{(1-n/2)n} \nonumber\\
g_s &=& \frac{(n-2d)^2}{(1-n/2)^2n^2}
\end{eqnarray}
The expression for $g_t$ is the same as in the early literature~\cite{brinkman}.
In the limit $d=0$, Eqns. (7) recover the previous results derived for the $t-J$ model
~\cite{zhang88}. These renormalized factors quantitatively describe the correlation effect
of the on-site repulsion. $g_t \leq 1$, and $g_t <<1$ at small $d$ and small $\delta$, 
representing the reduction of the kinetic energy due to the projection. 
$4 \geq g_s \geq 1$, and $g_s =4$ at
$d=0$ and $\delta=0$, representing the enhancement of the spin-spin correlation due to
the projection.
In Fig.2, we plot $g_t$ and $g_s$ as functions of the double occupation number $d$
for various electron densities.

\begin{figure}
\includegraphics*[width=2in,angle=270]{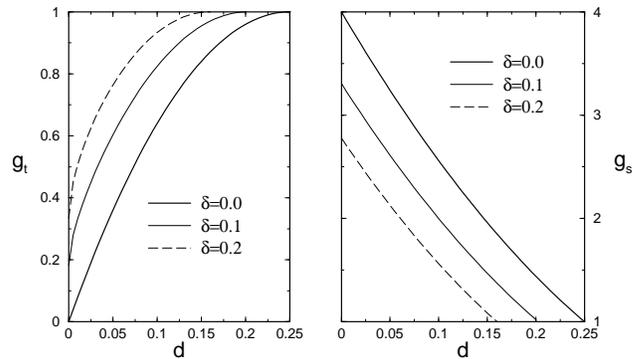}
\caption{The Gutzwiller's renormalization factors, $g_t$ and $g_s$,
as functions of the double occupation number $d$, obtained from Eqns. (6). $\delta = 1-n_e$. }
\label{fig:pd}
\end{figure}

In terms of these renormalization factors, we can define a renormalized Hamiltonian given by
\begin{eqnarray}
H' = g_t H_t + g_s H_s + H_U
\end{eqnarray}
The expectation value  of $H$ in the state $|\Psi_{GS}\rangle$ can be evaluated in terms of 
the expectation value of 
$H'$ in the state $|\Psi_{BCS}\rangle$. We obtain,
\begin{eqnarray}
E= \langle H' \rangle_0 = Ud+ g_t \langle H_t \rangle_0 + g_s\langle H_J \rangle _0
\end{eqnarray}
In the renormalized Hamiltonian approach, the original variational parameters $\{\alpha, v_{\vec k}, u_{\vec k}\}$
are transformed into the variational parameters $\{d, v_{\vec k}, u_{\vec k}\}$.  There is one to one correspondence
between $\alpha$ and $d$.  Within the Gutzwiller approximation, one can analytically
calculate $d = \langle n_{i\uparrow}n_{i\downarrow} \rangle$, and  one finds~\cite{gutzwiller,vallhardt}, 
\begin{eqnarray}
(1 - \alpha)^2 = \frac{d(1-n+d)}{(n/2-d)^2}
\end{eqnarray}

The energy in Eq. (9) is evaluated for the fixed number of electrons $N_e$. We introduce 
a Lagrangian multiplier $\tilde{\mu}$, and define
\begin{eqnarray}
K = H' - \tilde{\mu} (\sum_{i\sigma} n_{i\sigma} - N_e).
\end{eqnarray}
We then have $E= \langle K \rangle_0$, being subject to the condition $\partial \langle K \rangle_0 /\partial \tilde{\mu} =0$,
or
\begin{eqnarray}
2\sum_{\vec k} v_{\vec k}^2 =n
\end{eqnarray}
Below we consider the case $u_{\vec k}$ and $v_{\vec k}$ to be real.
Evaluating Eq. (11), we obtain (lattice constant =1),
\begin{eqnarray}
E & = & Ud + \tilde{\mu} + 2\sum\limits_{\vec{k}}(g_t\epsilon_{\vec{k}} -\tilde{\mu}) v_{\vec{k}}^2 \nonumber\\
  &   & + \sum\limits_{\vec{k},\vec{k^\prime}} V_{\vec{k}-\vec{k^\prime}}({v_{\vec{k}}}^2 {v_{\vec{k^\prime}}}^2 +
u_{\vec{k}}v_{\vec{k}}u_{\vec{k^\prime}}v_{\vec{k^\prime}})
\end{eqnarray}
where
\begin{eqnarray}
V_{\vec{k}} &=& -\frac{3}{2}g_sJ(\cos{k_x} + \cos{k_y}) \nonumber\\
\epsilon_{\vec{k}} &= &-2t(\cos{k_x} +\cos{k_y}) 
\end{eqnarray}
Carrying out the variational procedure with respect to 
$u_{\vec k}$ and $v_{\vec k}$,  we obtain
\begin{eqnarray}
u_{\vec{k}}^2 = \frac{1}{2}(1 + \chi_{\vec{k}}/E_{\vec{k}}) \nonumber\\
v_{\vec{k}}^2 = \frac{1}{2}(1 - \chi_{\vec{k}}/E_{\vec{k}})
\end{eqnarray}
where
\begin{eqnarray}
E_{\vec{k}} & = \sqrt{\chi_{\vec{k}}^2 + \Delta_{\vec{k}}^2}
\end{eqnarray}
The variational parameters $\Delta_{\vec{k}}$ and $\chi_{\vec{k}}$ are related to 
the particle-particle amplitude $\Delta_{\vec k}$ and the
particle-hole paring amplitudes $\chi_{\vec k}$ by,
\begin{eqnarray}
\Delta_{\vec{k}} &=&\Delta_x\cos{k_x}+ \Delta_y\cos{k_y}\nonumber \\
\chi_{\vec{k}} &=& \tilde{\epsilon_{\vec{k}}} - (\chi_x\cos{k_x}+ \chi_y\cos{k_y}).
\end{eqnarray}
In the above equations, we have introduced two correlation functions in the unprojected state
$|\Psi_{BCS} \rangle$,
\begin{eqnarray}
\begin{array}{l}
\Delta_{\tau}= \langle c_{i\downarrow}c_{i+ \tau,\uparrow} - c_{i\uparrow}c_{i+\tau,\downarrow} \rangle_0 \\
\chi_{\tau} = \sum\limits_\sigma \langle c_{i\sigma}^\dagger c_{i+ \tau \sigma} \rangle_0
\end{array}
\end{eqnarray}
with $\tau =x, y$, the unit vectors on the lattice,  and 
\begin{eqnarray}
\tilde{\epsilon_{\vec{k}}} & = & [-2g_t t(\cos{k_x} + \cos{k_y}) - \tilde{\mu}]/(3g_sJ/4).
\end{eqnarray}
For the $d$-wave pairing state, which has the lowest energy within this class of states
as suggested in the previous studies for the $t-J$ model~\cite{zhang88,kotliar,fukuyama,zhang91},
we have $\Delta_x = -\Delta_y = \Delta$, and 
$\chi_x=\chi_y=\chi$. $\Delta$ and $\chi$ are determined by the
coupled gap equations,
\begin{eqnarray}
\Delta &=& \sum\limits_{\vec{k}} (\cos{k_x})\Delta_{\vec{k}}/E_{\vec{k}} \nonumber\\
\chi &=& - \sum\limits_{\vec{k}} (\cos{k_x})\chi_{\vec{k}}/E_{\vec{k}}
\end{eqnarray}
These gap equations must be solved simultaneously with the hole concentration equation,
Eq. (12), which can be rewritten as  
$\delta=\sum\limits_{\vec{k}}\chi_{\vec{k}}/E_{\vec{k}}$, with $\delta =1-n_e$. 
The variation with respect to $d$ leads to the equation
\begin{eqnarray}
\frac{\partial E}{\partial d}= U + \frac{\partial g_t}{\partial d} \langle H_t \rangle_0 +
\frac{\partial g_s}{\partial d} \langle H_J \rangle_0 = 0.
\end{eqnarray}
In terms of $\chi$ and $\Delta$, the energy is given by
\begin{eqnarray}
E = Ud - 4n g_t t \chi - (3g_sJ/4)(\Delta^2 + \chi^2)
\end{eqnarray}
where $\chi$ and $\Delta$ are the solutions of the gap equations, and both are functions of $d$.
In the case there are multiple solutions for $d$ from Eq. (21),
the ground state is determined by the global energy minimum.
Alternatively, we may solve the gap equations 
for given values of $d$, and calculate $E(d)$ to find the optimal value of $d$ to determine the
ground state and the ground state energy.  The chemical potential of the system, 
$\mu = \partial E/\partial n$, is given by 
\begin{eqnarray}
\mu & = & \tilde{\mu} + \frac{\partial{g_t}}{\partial{n}} \langle H_t \rangle_0
                      + \frac{\partial{g_s}}{\partial{n}} \langle H_J \rangle_0
\end{eqnarray}
Note that chemical potential here is different from the Lagrangian multiplier $\tilde \mu$
in the renormalized mean field theory. This is because the renormalized factors $g_t$, $g_s$
to be also functions of electron density $n$.

\section*{\bf 4. Mott Insulator-Gossamer Superconductor Transition}

In this section, we discuss the variational solutions at the half filled case.
At the half filling, the trial wavefunction $|\Psi_{GS} \rangle$ describes either a Mott insulator
if $\alpha=1$ (i.e. $d=0$), or a superconducting state if $\alpha <1$
(i.e. $d>0$).  If $\alpha$ is close to $1$, or $d$ is very close to zero, $|\Psi_{GS} \rangle$ describes a
gossamer superconducting state.

We expect a Mott insulator at large $U$ and a superconducting state at small $U$. This can be
examined qualitatively without carrying out the quantitative calculations.
At the half filling, $g_t = 8(1-2d)d$, and $g_s = 4(1-2d)^2$.
Eq. (21) becomes
\begin{eqnarray}
U + 8(1-4d) \langle H_t \rangle_0 - 16(1-2d) \langle H_J \rangle_0 =0.
\end{eqnarray}
Since both $\langle H_t \rangle_0$ and $\langle H_J \rangle_0$ are finite, there will be no
solution of Eq. (24) if $U$ is sufficiently large. This indicates that the ground state
corresponds either $d=0$ or $d=d_{max}$, the allowed maximum value of $d$. The repulsive nature of $U$
excludes the latter, and it follows that the Mott insulating
state with $d=0$ is the ground state. We believe that the qualitative result for the existence of the
Mott insulating phase at large but finite $U$
is robust.  Note that in the Gutzwiller's wavefunction, the doubly occupied site and the empty site
are not correlated. At the half filling, $d$ represents the carrier density $n^*$
and is proportional to the Drude weight in the a.c. conductivity, $n^*e^2/m^*$, with $m^*$ the effective mass.
We remark that the parameter $d$ in our Gutzwiller
approach is different from the usual double occupation number
$\tilde d$ (for example, the double occupation calculated in the exact diagonalization
 of a finite size system). In the latter
case, $\tilde d$ also includes the contribution from the virtual hopping process. Therefore,
the double occupied site is bound to the empty site, and the double occupation number
$\tilde d$ does not represent the mobile carriers.

In the insulating phase, the ground state is the same as that of the Heisenberg model.
Within our theory, the ground state energy is given by
\begin{eqnarray}
E_0 = -3J (\Delta_0^2 + \chi_0^2)
\end{eqnarray}
with
\begin{eqnarray}
\Delta_0 &=&\chi_0 =C/\sqrt 2 \nonumber\\
&=&\frac{1}{\sqrt 8}\sum_{\vec k}\sqrt{\cos^2{k_x} + \cos^2{k_y}} =0.339.
\end{eqnarray}
At small $U$, one expects a metallic ground state, except in the special cases due to the band effect,
such as having the Von Hove singularity and
perfect nesting in $H_t$. In this paper, we will not consider the special band effect. Therefore,
we expect a metal-insulator transition at a finite $U=U_c$ in the general case,
with the metallic phase to be superconducting provided that $u_{\vec k}v_{\vec k} \neq 0$.
This is the Mott
insulator-gossamer superconductor transition.

We now discuss the phase transition in details.  We solve the gap equations for the fixed $d$ and determine
the transition point $U_c$ and the nature of the transition.  The phase diagram in the parameter space
$U$ and $J/t$ is plotted in Fig. 3.  The critical $U_c$ separates the Mott insulating phase from
the gossamer superconducting phase. We can choose the
mobile carrier density as the order parameter, which is proportional to $d$.
The phase transition is classified as the second type if $d \rightarrow 0$ and
the first type if $d \rightarrow d_c >0$
as $U \rightarrow U_c$ within the metallic phase.
This classification is consistent with the usual
zero temperature quantum phase transition, where the
nature of the phase transition depends on the continuity or discontinuity
of the order parameter.  We find
the transition to be the first order at $0 < J/t < \eta_c$, and the second order at $\eta_c < J/t$,
with $\eta_c \approx 2$. At $J=0$, the present theory is reduced to the Brinkman-Rice theory
for metal-insulator transition~\cite{brinkman}
for the projected non-interacting electron state.
In that case, we find $U_c/t = 128/\pi^2$.  From Fig. 3,  we see that
$U_c (J \rightarrow 0) = U_c (J =0)$, so that the critical value of $U$ is continuous at $J=0$.
However, the transition is the second type at $J=0$, while it is
the first type for any small but finite $J/t$.

\begin{figure}
\includegraphics*[width=4.5cm,height=6.5cm,angle=270]{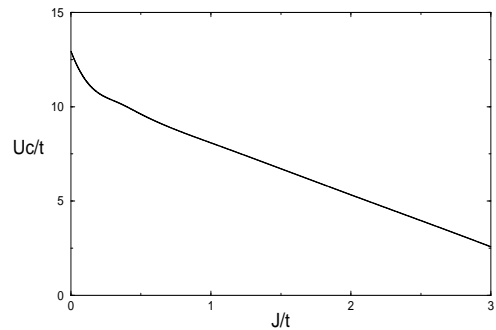}
\caption{Phase diagram at the half filling. }
\label{fig:pd}
\end{figure}

Let us first discuss the first order phase transition in the region $0 < J/t < 2$. In Fig. 4,
we show the energy $E$ as a function of $d$ for several values of $U$ at a typical parameter $J/t=1/3$.
$E$ is not a monotonic function of $d$ for $U$ near $U_c$.
As $d$ increases from $d=0$, $E$ first increases linearly, then decreases, then increases again.
There is a local energy minimum around $d=0.02$, which  develops and becomes a global minimum
as $U$ approaches $U_c$ from the insulator side.
The local minimum $E(d_c)$ at $d=d_c$ represents a metallic solution, and $E(d=0)$
represents an insulator solution. The critical value for the Mott insulator
and gossamer superconductor transition is determined by the condition $E(d_c)= E(d=0)$.
From Fig. 4, we have
$U_c/t=10.23$ for $J/t=1/3$.  At $U>U_c$, $d=0$, and the ground state is an insulator.  At $U < U_c$,
$d \geq d_c \approx 0.02$, and the ground state is
a gossamer superconducting state.  In Fig. 5, we plot $d$ as a function of $U$.
$d$ is approximately linear in $U$
till the transition point $U_c$.  The discontinuity in $d$ is about 0.02.
We conclude that the Mott insulator-gossamer superconductor phase 
transition in this most relevant region is first type.
The carrier density is discontinuous at the phase transition point.  Since $d$ is proportional to the
carrier density, this type of first order transition
should be observable in the electric transport or in the a.c. conductivity measurements.

\begin{figure}
\includegraphics*[width=4.5cm,height=6.5cm,angle=270]{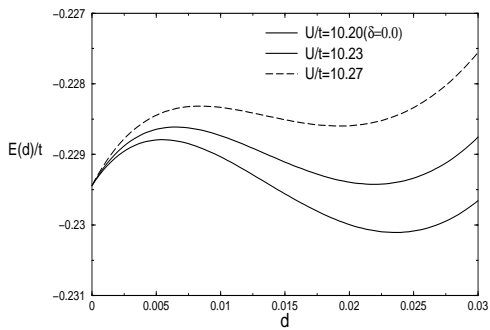}
\caption{The energy E as a function of $d$ for
$U$ around
the critical value $U_c$ at $\delta=0$. The ratio $J/t=1/3$.}
\label{fig:pd}
\end{figure}

\begin{figure}
\includegraphics*[width=4.5cm,height=6cm,angle=270]{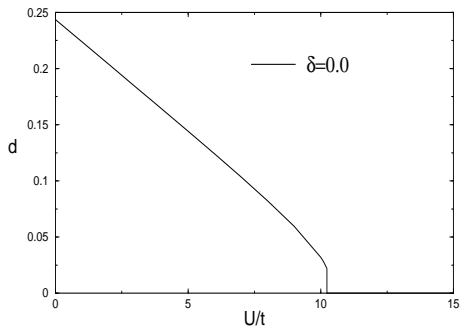}
\caption{The double occupation number $d$ as a function of $U$ at $\delta=0$ and $J/t$=1/3. }
\label{fig:pd}
\end{figure}

For large ratio of $J/t$, our calculations show that the phase transition is second order.
This is illustrated in Fig. 6 for $E$ v.s. $d$ in the case of $J/t=3$.  The transition
occurs at $U_c =2.58t$, and $d$ changes continuously across $U_c$. 

\begin{figure}
\includegraphics*[width=4.5cm,height=6cm,angle=270]{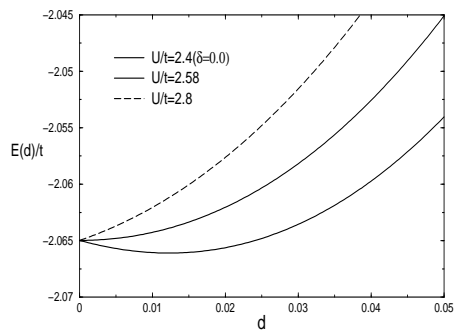}
\caption{The energy E as a function of the double occupation number d for
several values of $U$ around
the critical value $U_c$ with $\delta$=0 and $J/t$=3.}
\label{fig:pd}
\end{figure}

A special case is $J=0$. In this limit, $|\Psi_{GS} \rangle = \Pi_{\alpha}|\Psi_{FL} \rangle$, and 
our theory is reduced to the previous one for the projected Fermi liquid state. 
The energy in Eq. (9) at the half filling becomes
\begin{eqnarray}
E &=& Ud - 2g_t\sum_{\vec k} |\cos{k_x} + \cos{k_y}| \nonumber\\
&=& Ud -128d(1-2d)t/\pi^2
\end{eqnarray}
From this we find $U_c/t = 128/\pi^2 \approx 13$. $d$ is 
continuous at $U_c$ so that the transition is second type.

Our result on the first order phase transition in the physically interesting region (small but non-zero $J/t$)
is somewhat unexpected.
We argue that the first order transition between the Mott insulator and the gossamer superconductor
is due to the interplay 
of the kinetic and spin-spin correlation energies. This interplay 
was not included in the previous study of the Gutzwiller approach but
is taken into account here. To illustrate the effect of the interplay 
to the nature of the phase transition, we consider the limiting case $0< J/t << 1$, 
and expand the energy $E$ of Eq. (22) at $n=1$ for small $d$,
\begin{eqnarray}
E(d) = E_0 + (U - U_{c0})d - \beta d^2 + O(d^3)
\end{eqnarray}
where $E_0$ is the energy at $d=0$ given by Eq. (25), 
$U_{c0}= 16\sqrt{2}Ct - 12C^2J$ is the solution of $\partial E/\partial d =0$ at $d=0$, given by
Eq. (24).
$\beta = [ 32 (\partial \chi/\partial d)_{|_{d=0}} - 32 \sqrt{2}C) ]t$. The $J$-dependence in $\beta$ has been
neglected since $J/t <<1$.  Note that the kinetic energy is proportional to $\chi$.
As $d$ increases from $0$, $\chi$ tends to increases from $\chi_0 = C/\sqrt{2}$
to gain more kinetic energy. Therefore, $\partial{\chi}/\partial d >0$. In the limit $J/t <<1$, we have
$\partial \chi/\partial d =(\chi - \chi_0)/d \propto t/J >>1$, hence the first term in the expression for
$\beta$ dominates
and $\beta > 0$.  This demonstrates that $d=0$ is a local maximum in energy at $U= U_{c0}$, and the
phase transition 
occurs at a large value of $U$ corresponding to $d >0$ as numerically shown in Fig. 4, hence it is a first order transition.
Numerically, we find that $\beta = 34.8472$, for $J/t=1/3$.

It is interesting to compare the gossamer superconductor - Mott insulator transition with the metal-insulator
transition
studied in previous literature~\cite{brinkman}.  In the Brinkman-Rice theory, the transition is second order.
In that theory, the system approaches the insulating phase,
the effective mass $m^* \rightarrow \infty$. In the gossamer
superconductor-Mott insulator transition with small ratio of $J/t$, the insulating phase is not characterized by the
divergence of the effective mass. We estimate the ratio of the effective mass to the band mass (1/t)
at the metallic side of transition point 
to be $1/g_t \approx 1/(8d)\approx 6$.

The first order phase transition between metal and insulator was first pointed out by
Peierls~\cite{peiperls} and by Landau and Zeldovich~\cite{landau}, and examined
in more great detail by Mott~\cite{mott}.
In their theory, an electron is always bound to a positive charge due to the long range
Coulomb attraction, and the transition of a metal to an insulator at zero or very low temperatures
occurs at a finite critical electron density, and must be first type.  
It is interesting to note that the on-site repulsion also leads to the first order transition
between a specific type of metal (superconductor) and an insulator studied in the present paper,
where the long range Coulomb force is not included. 
We also note that Florencio and Chao ~\cite{chao}
investigated the metal-insulator transition of the Hubbard model using Gutzwiller's
wavefunction by including antiferromagnetism and found the transition to be first type.

\section*{\bf 5. Gossamer and RVB superconductivity}

In this section, we discuss the superconducting state at the half filled as well as
away from the half filled.  Note that at $\delta > 0$, $|\Psi_{GS} \rangle$ always describes
a metallic state.  To make the terminology clearly, we shall call the superconducting state
at $U < U_c$ to be the gossamer superconductor~\cite{laughlin,zhang02}, and the
doped Mott insulator ($U > U_c$ and $\delta > 0$) to be the RVB state~\cite{anderson}.

We begin with the discussion of the double occupation number $d$ as a function of
the hole concentration. We solve the gap equations and find the optimal value of $d$.
The results are plotted in Fig. 7. We find that $d$ is always non-zero at $\delta > 0$, even
in the region $U > U_c$. This suggests that the doped Mott insulator is described by
a partially projected state ($\alpha <1$ in Eq. (2)).  Nevertheless, $d$ is very small for $U \gg U_c$.
As we can see from Fig. 7, $d$ varies from $0$ to $0.01$ for $U/t =15$, which corresponds to
$U/U_c \approx 1.5$. The non-zero value of $d$ at $\delta > 0$ may be understood from 
the variational equation (21), which determines $d$. 
At $\delta >0$, $\partial g_t/\partial d_{|_{d=0}} \rightarrow \infty$. Therefore, $d=0$ cannot be
a solution of the equation, and $d$ must be finite.  It remains to be seen
if this result is due to the Gutzwiller's approximation used in our calculation.
It will be interesting to further
examine this issue using other methods such as the variational Monte Carlo method. 

From Fig. 7, we also see that as $\delta$ increases from $0$, $d$ increases for large $U$ while
$d$ decreases for small $U$. The latter may be understood as follows.
In the small $U$ case, the correlation becomes less important, and
the qualitative feature between $d$ and $\delta$ becomes similar to the uncorrelated state.
For the uncorrelated Fermi liquid state, $d=(1-\delta)^2/4$, so that $d$
monotonically decreases as $\delta$
increases.

\begin{figure}
\includegraphics*[width=4.5cm,height=5.8cm,angle=270]{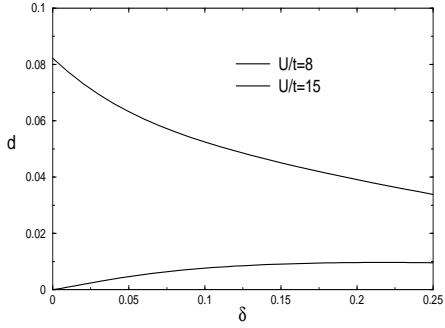}
\caption{The double occupation number $d$ as a function of
$\delta$ at the ratio $J/t=1/3$ for several $U$'s. }
\label{fig:pd}
\end{figure}

While $d$ is a smooth function of $\delta$ for most values of $U$ in our study, there is a narrow
region in $U$ above $U_c$, $d$ changes discontinuously at a very small $\delta$.
In Fig. 8, we show the energy $E$ v.s. $d$ for $U=10.235t$, slightly above $U_c=10.23t$,
for four values of $\delta$.  At $\delta=0$, $d=0$ corresponds to the global energy minimum,
while there is a local minimum around $d=0.02$. As $\delta$ gradually increases, the positions of the
two minima change smoothly and their corresponding energies reverse their order. In this region,
the optimal value of $d$ jumps.  This region is found very narrow:  $10.23t < U < 10.235t$, however.

\begin{figure}
\includegraphics*[width=8cm,height=8cm,angle=270]{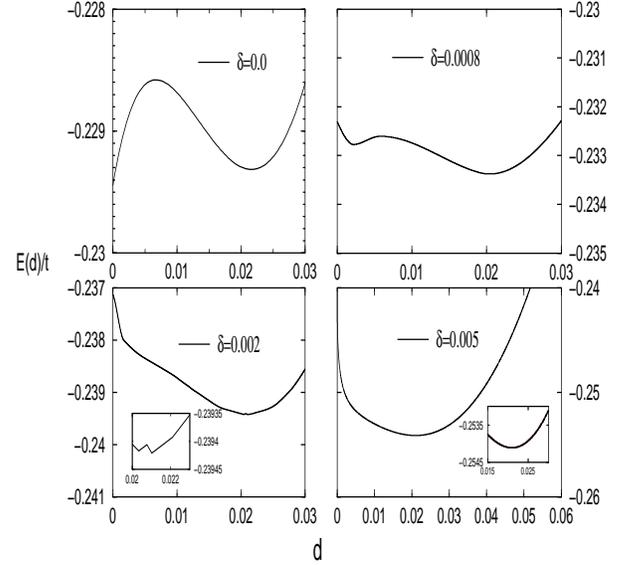}
\caption{Energy $E$ as a function of $d$ for $U=10.235t \gtrsim U_c$ at the ratio $J/t=1/3$, for several values
of $\delta$.}
\label{fig:pd}
\end{figure}

We now discuss the superconducting order parameter. The superconducting order parameter of the
state $|\Psi_{GS}\rangle$ is defined by, for the d-wave pairing,
\begin{eqnarray}
\Delta_{sc}(\tau) = \langle c_{i\downarrow}c_{i+ \tau \uparrow} \rangle 
 -\langle c_{i\uparrow}c_{i+ \tau\downarrow}\rangle,
\end{eqnarray}
and $\Delta_{sc}=\Delta_{sc}(x) = - \Delta_{sc}(y)$. We shall adopt the Gutzwiller approximation to
calculate this quantity. In analogy to the derivation for the hopping energy in Eq.(6),
we find that~\cite{zhang88}
\begin{eqnarray}
<c_{i\downarrow}c_{i+\tau\uparrow}> & = & g_t <c_{i\downarrow}c_{i+\tau\uparrow}>_0
\end{eqnarray}
Therefore the order parameter $\Delta_{sc}$ is related to the variational parameter
$\Delta$ in the gap equations by
\begin{eqnarray}
\Delta_{sc} & = & g_t\Delta
\end{eqnarray}
In Fig. 9, we show our results for
$\Delta_{sc}$ and $\Delta$ as functions of $\delta$ for three values of $U$: well above $U_c$,
at $U_c$, and well below $U_c$.  Note that at $U=15t >>U_c$, $\Delta_{sc} =0$ at $\delta =0$, 
although $\Delta$ takes a maximum.  This is consistent with the Mott insulating ground state.
At $\delta$ increases, the kinetic energy plays more important role in comparison with the spin-spin
correlation energy, and $\Delta$ decreases monotopically. However, $\Delta_{sc}$ shows a 
non-monotonic dome
shape,  and it first  increases
to reach a peak before it drops for larger $U$. Also note that at $U_c$ the Mott insulator and gossamer
superconducting state are degenerate at $\delta=0$, and the gossamer superconducting phase continuously
evolves into the metallic phase at $\delta >0$. Shown in the figure for $U=10.23t \approx U_c$
is the metallic phase. The non-zero value of $\Delta_{sc}$ at $\delta=0$ 
indicates the transition to be the first order.

\begin{figure}
\includegraphics*[width=3in,height=2.5in,angle=270]{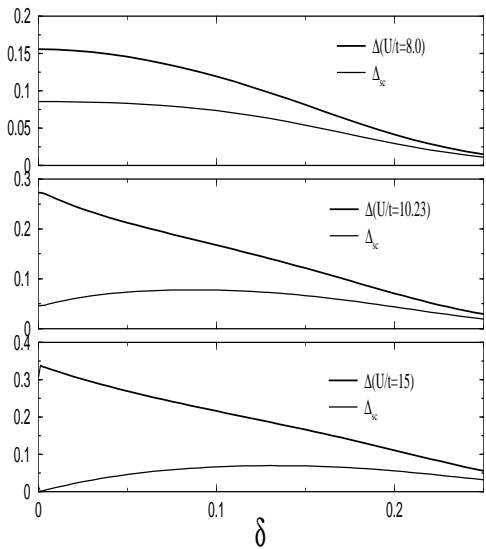}
\caption{Variational parameter $\Delta$ and superconducting order parameter $\Delta_{sc}$ as
functions of the hole concentration $\delta$ for three values of $U$ at ratio $J/t$=1/3.}
\label{fig:pd}
\end{figure}

In Fig. 10,we show $\Delta_{sc}$ and $\Delta$ as a function of $U$ for $\delta=0$.
From this figure we can see that for $U<U_c$, $\Delta_{sc}$ and $\Delta$ both increase with $U$.
At $U\ge U_c$ $\Delta$ takes a maximum while $\Delta_{sc}=0$. Note that at small values of $U$,
the effective Hamiltonian does not represent the original Hubbard model, and one should be cautious to
interpret the results at small $U$.
\begin{figure}
\includegraphics*[width=3in,height=2.5in,angle=270]{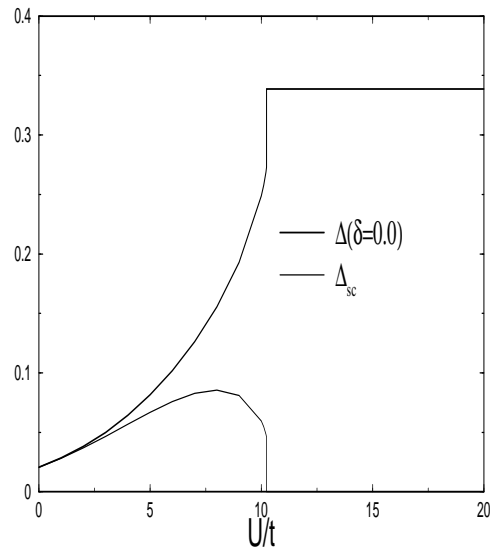}
\caption{Variational parameter $\Delta$ and superconducting order parameter $\Delta_{sc}$ as
functions of $U$ at $\delta=0$ and $J/t=1/3$.}
\label{fig:pd}
\end{figure}

It is interesting to point out that the superconducting order parameter $\Delta_{sc}$
in both the gossamer and  RVB superconducting states are characterized by the
variational parameter $\Delta$ and a
small renormalized factor $g_t$.  They have the same pairing symmetry, and the two states are
adiabatically connected to each other. From this point of view, the gossamer 
and RVB superconductors are the same.  In Fig. 11, we present a schematical diagram for the
Mott insulator, gossamer and RVB superconducting states. In the parameter space of Coulomb
repulsion $U$ and of the hole concentration $\delta$, there is a line at $\delta=0$ and at $U > U_c$
for the Mott insulating phase.  In the region $U > U_c$ and $\delta >0$, it is the RVB superconducting phase
as Anderson proposed~\cite{anderson}.  In the region $U<U_c$, it is the gossamer superconducting phase
in which both the half filled and non-half filled
states are superconducting. 

\begin{figure}
\includegraphics*[width=6.5cm,height=5.5cm,angle=0]{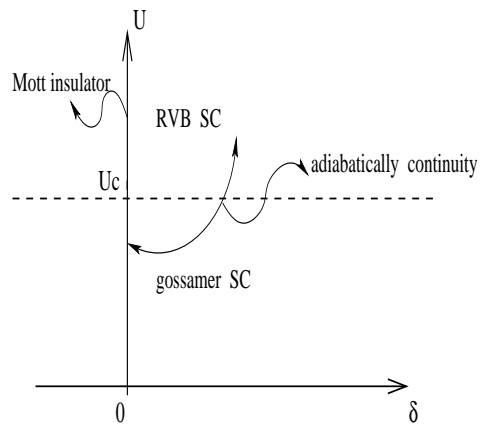}
\caption{Schematical phase diagram for the Hamiltonian in parameter space $U$ and $\delta$.}
\label{fig:pd}
\end{figure}

While the gossamer and RVB superconducting states are essentially the same, the chemical
potential $\mu$ in the gossamer superconducting state is continuous at $\delta=0$ because of the metallic 
phase, while $\mu$ is discontinuous at $\delta=0$ because the state at $\delta=0$ is an insulator and
the state at any small but finite $\delta$ is a metal within the present theory.

Below we shall study $\mu$ quantitatively.  At $\delta=0$, $\mu=U/2$ by electron-hole symmetry.
At other value of $\delta$, we calculate $\mu$ using Eq. (23) after solving the gap equations.
In Fig. 12, we show $\mu$ as a function of $\delta$.  As we can see from the figure, $\mu=U/2$ at
$\delta=0$, and is continuous for $U \leq U_c=10.23t$.  There is a discontinuity in $\mu$ for
$U >U_c$ at $\delta=0$.  At $U>U_c$, the chemical potential is shifted from $U/2$ at 
the half filled to the lower Hubbard band away
from the half filled. To see this more explicitly,
we define $\Delta\mu = \mu(d\rightarrow 0) - \mu(d=0)$.
$\Delta\mu$ as a function of $U$ is plotted in
Fig. 13.  As $U$ decreases, $\Delta \mu$ decreases monotonically and reaches a finite value
at $U=U_c+0^+ $, then drops to zero at $U=U_c -0^+$. The discontinuity of $\Delta\mu$ at $U_c$
is related to the first order phase transition.

\begin{figure}
\includegraphics*[width=4.5cm,height=5.8cm,angle=270]{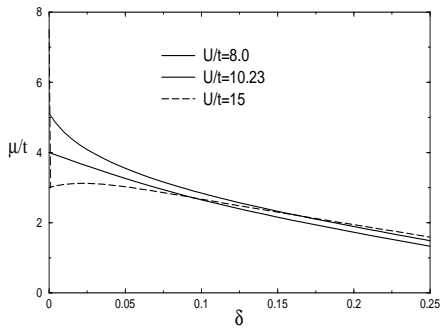}
\caption{The chemical potential $\mu$ as a function of the hole concentration
$\delta$ for several values of intra-site Coulomb repulsion $U$ with $J/t=1/3$.}
\label{fig:pd}
\end{figure}

\begin{figure}
\includegraphics*[width=4.5cm,height=5.5cm,angle=270]{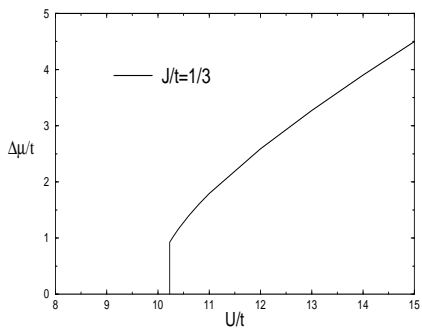}
\caption{The discontinuous in the chemical potential as a function of intra-site Coulomb
repulsion $U$ for $J/t$=1/3.}
\label{fig:pd}
\end{figure}

Finally, we briefly discuss the excited states. 
In the context of the Gutzwiller trial wavefunction, the excited states were discussed
by Zhang et al. ~\cite{zhang88} for the RVB state, and recently by Laughlin for a 
gossamer superconductor Hamiltonian\cite{laughlin}. Following Zhang et al., the quasiparticle
states can be defined by
\begin{eqnarray}
|\Psi_{\vec p \uparrow}\rangle =\Pi_{\alpha}c^{\dagger}_{\vec p \uparrow}\prod_{\vec k \neq \vec p}
(u_{\vec k} +v_{\vec k}c^{\dagger}_{\vec k\uparrow}c^{\dagger}_{-\vec k\downarrow})|0\rangle
\end{eqnarray}  
The quasiparticle energy $\tilde E_{\vec p}$ \,
is defined to be the difference of the expectation values of $K$ in Eq. (32)
in this state and in the ground state $|\Psi_{GS}\rangle$. We use the Gutzwiller method to calculate
the energy and obtain~\cite{zhang88}
\begin{eqnarray}
\tilde E_{\vec p} = (3g_sJ/4)\sqrt{\chi_{\vec p}^2 + \Delta_{\vec p}^2}.
\end{eqnarray}
At the vectors $\vec p$ satisfying $\chi_{\vec p} =0$ (Eq. (17)), we have 
$\tilde E_{\vec p} = \Delta |\cos{k_x} - \cos{k_y}|$.
Therefore, the quasipartical energy is proportional to the parameter $\Delta$,
and not renormalized by the factor $g_t$, which is very different from the superconducting
order parameter. From Fig. 9, our theory
predicts the quasipartical energy to be maximum at $\delta=0$, and decreases as doping increases.
This feature was first found for the large-U limit Hubbard model~\cite{zhang88}, and
is consistent with the "high energy pseudogap" seen in the angular resolved photoemission
experiments and the experimentally observed superconducting energy gap~\cite{norman,shen,campuzano}.
Here we show that this feature appears also in the gossamer superconductor.

\bigskip
\section*{\bf 6. Summary}

We have used the Gutzwiller variational method to study an effective Hamiltonian for 2-dimension
Hubbard model.Based on Gutzwiller approximation, we have discussed the case both at the half filled
and away from the half filled. At the half filled, there is a first order phase transition to separate a Mott insulator
at large Coulomb repulsion $U$ from a gossamer superconductor at small $U$.This is very interesting.
It suggests that the on-site Coulomb repulsion can lead to the first transition between a specific
type of metal and an insulator.The double occupation number d which is proportional to the carrier
density changes discontinuously from zero in the
Mott insulator phase to a finite value at the phase transition
point($U=U_c$). So we expect that this type of first order transition should be observable in the electric
transport or in the a.c. conductivity measurements. Away from the half filled, the Gutzwiller variational state is
always metallic. The gossamer superconducting state changes continuously from
its state at the half filled, while the Mott insulating phase becomes RVB superconducting.
The gossamer superconductor is similar to the RVB suerconducting states with the same paring symmetry, and 
showing the pseudogap. Their major difference is on the position of their chemical potential.
The Gutzwiller method we used in this paper has previously been tested in good agreement
with variational Monte Carlo method~\cite{zhang88,gros}. We believe that the qualitative
conclusions obtained here should be reliable, and refined numerical calculations such as variational Monte Carlo
calculations will be interesting to examine the problem.

There are other questions that require further investigation such as the effect of the
antiferromagnetism which is most plausible in this model. 

We wish to thank R. Laughlin, T. M. Rice, P. W. Anderson, Y. Yamashita, and Y. Yu for many interesting discussions.
This work was partially supported by the US NSF grant 0113574, and by 
Chinese Academy of Sciences.

\end{document}